# Focusing of in-plane hyperbolic polaritons in van der Waals crystals with tailored infrared nanoantennas

Javier Martín-Sánchez[1,2] [*&], Jiahua Duan[1,2&], Javier Taboada-Gutiérrez[1,2], Gonzalo Álvarez-Pérez[1,2], Kirill V. Voronin[3], Iván Prieto[4], Weiliang Ma[5], Qiaoliang Bao[6], Valentyn S. Volkov[3,7], Rainer Hillenbrand[8,9], Alexey Y. Nikitin[9,10], Pablo Alonso-González [1,2 *]

[1] *Department of Physics, University of Oviedo, Oviedo, Spain*
[2] *Center of Research on Nanomaterials and Nanotechnology, CINN (CSIC—University of Oviedo), El Entrego 33940, Spain*
[3] *Center for Photonics and 2D Materials, Moscow Institute of Physics and Technology, Dolgoprudny 141700, Russia*
[4] *Institute of Science and Technology Austria IST, Am Campus 1, 3400 Klosterneuburg, Austria*
[5] *Wuhan National Laboratory for Optoelectronics, School of Optical and Electronic Information, Huazhong University of Science and Technology, Wuhan, China*
[6] *Department of Applied Physics, The Hong Kong Polytechnic University, Hung Hom, Kowloon, Hong Kong, P. R. China*
[7] *GrapheneTek, Skolkovo Innovation Center, Moscow 143026, Russia*
[8] *CIC nanoGUNE, Donostia-San Sebastián, Spain*
[9] *IKERBASQUE, Basque Foundation for Science, Bilbao, Spain*
[10] *Donostia International Physics Center (DIPC), Donostia-San Sebastián, Spain*

[&] These authors contributed equally to this work
[*] Corresponding author: javiermartin@uniovi.es, pabloalonso@uniovi.es

**Phonon polaritons (PhPs) – light coupled to lattice vibrations – with in-plane hyperbolic dispersion exhibit ray-like propagation with large wavevectors and enhanced density of optical states along certain directions on a surface. As such, they have raised a surge of interest as they promise unprecedented possibilities for the manipulation of infrared light with planar circuitry and at the nanoscale. Here, we demonstrate, for the first time, the focusing of in-plane hyperbolic PhPs propagating along thin slabs of α-MoO$_3$. To that end, we developed metallic nanoantennas of convex geometries for both the efficient launching and focusing of the polaritons. Remarkably, the foci obtained exhibit enhanced near-field confinement and absorption compared to foci produced by in-plane isotropic PhPs. More intriguingly, foci sizes as small as $\lambda_p/5 = \lambda_0/50$ were achieved ($\lambda_p$ is the polariton wavelength and $\lambda_0$ the photon wavelength). Focusing of in-plane hyperbolic polaritons introduces a first and most basic building block developing planar polariton optics utilizing in-plane anisotropic van der Waals materials and metasurfaces.**

Focusing of electromagnetic waves to deeply sub-diffractional volumes allows for enhancing light-matter interactions, which is the key for a broad range of applications at the nanoscale such as light manipulation [1–3], nanolithography [4,5], photocatalysis [6,7], and bio-sensing [8–10]. Recently, hyperbolic PhPs [11] in thin slabs of the vdW crystal h-BN have been intensively investigated as they exhibit extreme field confinement [12,13] and exotic ray-like propagation [14,15], with potential for hyper-lensing [16] and focusing [15,17] of mid-infrared light at the nanoscale. However, hyperbolic PhPs in h-BN exhibit out-of-plane hyperbolic propagation, challenging the development of hyperbolic nanooptics compatible with on-chip optical devices [18]. In this regard, hyperbolic PhPs with in-plane propagation have recently been demonstrated in artificial h-BN metasurfaces [19] and in the natural vdW crystals α-V$_2$O$_5$ [20] and α-MoO$_3$ [21], which allows fundamental and applied studies of hyperbolic focusing phenomena in a planar configuration [18].

Here, we develop metallic Au nanoantennas with tailored geometries to excite and focus in-plane hyperbolic PhPs along the surface of a α-MoO$_3$ slab. Our theoretical and experimental findings show the possibility of obtaining extraordinarily small foci with enhanced near-field confinement and absorption in hyperbolic media with respect to isotropic media. This result can be explained by the interference of highly directional hyperbolic polaritons featuring large wavevectors and enhanced density of electromagnetic modes.

**Results**

**In-plane propagation and focusing of hyperbolic PhPs in α-MoO$_3$.** The in-plane hyperbolic propagation of PhPs arises from a different sign of the permittivity of the host material along the two in-plane crystalline directions. It can be described by a hyperbolic iso-frequency curve (IFC) - a slice of the polariton dispersion in the momentum-frequency space defined by a plane of constant frequency (ω) –, as shown in Fig. 1a for PhPs in a α-MoO$_3$ slab at infrared frequencies (illuminating wavelength $\lambda_0$=11.05 μm). According to this IFC, the propagation of PhPs in α-MoO$_3$ is only allowed along specific directions laying within the sectors $|\tan(k_x/k_y)| < \sqrt{-\varepsilon_y/\varepsilon_x}$ between the asymptotes of the hyperbola in the ($k_x$, $k_y$) space (x and y corresponding to the α-MoO$_3$ [001] and [100] crystalline directions, respectively). Furthermore, the Poynting vector $\vec{S}$, which determines the propagation direction of PhPs in real space and is perpendicular to the IFC, is generally non-collinear with the wavevector $\vec{k}$. This is in stark contrast to the propagation of waves in isotropic media where $\vec{S}$ and $\vec{k}$ are collinear, and thus leads to exotic and non-intuitive optical phenomena. Remarkably, when approaching the two asymptotes of the IFC, the number of available wavevectors of PhPs largely increases (high-$|\vec{k}|$ wavevectors, denoted by $\vec{k}_H$), which yields a highly directional ray-like propagation as shown in Fig. 1a [15,16].

To further understand the ray-like character of in-plane hyperbolic propagation of PhPs in α-MoO$_3$, it is useful to analyze the density of electromagnetic modes (color-plot in Fig. 1b), which is proportional to the Fourier image of the vertical electric field generated by a vertical point source (see details in Supplementary Information S1). One can observe that the maxima are obtained for directions closely aligned with the asymptotes of the IFC, i.e., for PhPs with wavevectors $\vec{k}_H$ and Poynting vector $\vec{S}$ as indicated in Fig. 1a. As a result, the pattern of the field emitted by the vertical point source exhibits very narrow lobes along directions forming an angle θ=20º with respect to the y-axis (shown by dashed yellow lines in Fig. 1b). Note that such directional propagation can be explained from simple geometrical considerations: the Poynting vector $\vec{S}$ is perpendicular to the IFCs and thus to the maxima of the density of allowed modes in the region of large $\vec{k}$ wavevectors. These analytical results prove, therefore, that the ray-like propagation of PhPs along slabs of α-MoO$_3$ is consistent with the high density of electromagnetic modes associated with the asymptotes of the IFCs. Additionally, the latter leads to the existence of concave wavefronts centered along the y-axis, as demonstrated in Fig. 1c, where the real part of the full-wave simulated near-field distribution, $Re(E_z(x, y))$, of a vertical point dipole in close proximity to α-MoO$_3$ is represented (PhPs with $\vec{k}_H$ wavevectors are marked with black dashed arrows). For comparison, the case of PhPs propagating in a conventional isotropic medium is shown in the inset to Fig. 1c, where divergent PhPs propagation with convex wavefronts is instead observed.

In the following, we study the focusing of in-plane hyperbolic PhPs in α-MoO$_3$. To this aim, we propose the use of optical nanoantennas, which would allow for both effective launching and eventual focusing of the PhPs. Figure 1d shows the near field, $Re(E_z(x, y))$, calculated by full-wave numerical simulations for a metal nanoantenna with in-plane circular geometry (disk) located on top of a α-MoO$_3$ slab at $\lambda_0$=11.05 μm. We observe PhPs launched by the nanoantenna that exhibit large near-field amplitude and convex wavefronts within a triangular region, in which a focusing spot is formed at its apex (marked with a yellow dashed circle). Excitingly, by plotting a profile of the electric field intensity, $|E_z(x, y)|^2$, along the x-axis in Fig. 1d (red dashed line), we obtain a full width at half maximum (FWHM) of the spot of ~390 nm (red line in Fig. 1e), revealing a deep-subwavelength size of ~$\lambda_0/28$. Moreover, when comparing this profile with that obtained at the same distance from a point source launching PhPs in α-MoO$_3$ (blue line in Fig. 1e corresponding to the blue dashed line in Fig. 1c), we observe a field enhancement of about x12.

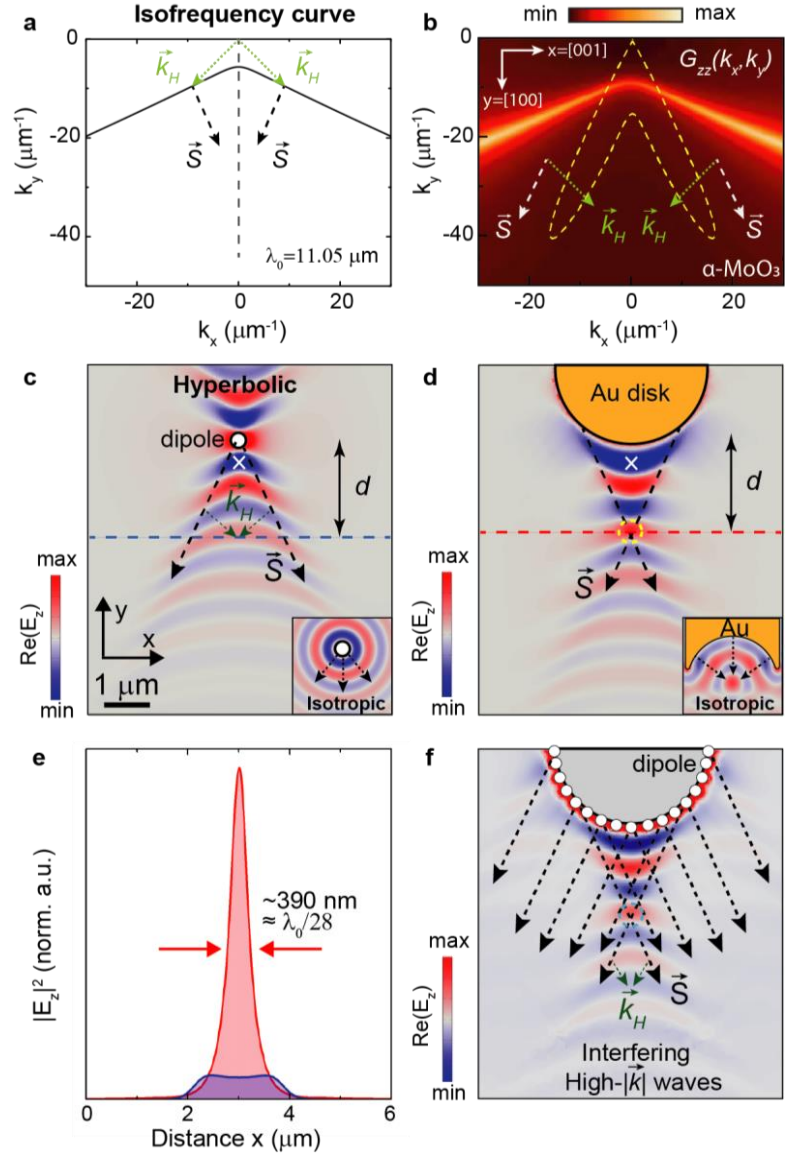

**Figure 1.- Propagation and focusing of in-plane hyperbolic PhPs. a,** IFC of hyperbolic PhPs in a 165-nm-thick α-MoO$_3$ crystal at an illuminating wavelength $\lambda_0$=11.05 μm. PhPs with high-$|\vec{k}|$ wavevectors propagating closely along the asymptote of the hyperbola are indicated by $\vec{k}_H$, together with their Poynting vector $\vec{S}$. **b,** The dashed yellow line represents the pattern of the field emitted by the vertical point source as a function of the polar angle. The color-plot corresponds to the density of electromagnetic modes in the k-momentum space ($k_x,k_y$). **c,** Simulated real part of the near-field distribution, *Re(Ez)*, of propagating PhPs excited by a vertically-oriented electric point dipole situated over the surface of the α-MoO$_3$ crystal. The inset shows the case for an isotropic medium. **d,** Same than (c) for a metal Au disk nanoantenna: the interference of PhPs with $\vec{k}_H$ wavevectors leads to a single focal spot (yellow dashed circle). The inset shows the case for an isotropic medium when a rod-like metal antenna with a concave extremity is used as excitation source. **e,** Profiles of the electric field amplitude, $|E_z|^2$, along the x-axis ([001] crystal direction) at the positions indicated by a blue-red dashed line in (c)-(d), respectively. A deep sub-wavelength focal spot with FWHM~390 nm ($\lambda_0/28$) is revealed when a disk-like nanoantenna is employed. The near-field intensities are normalized to the intensity of the first interference fringe marked with a white cross in (c)

and (d). **f,** Simulated near-field distribution, *Re(E$_z$)*, for a discrete distribution of point electric dipoles localized along the periphery of a virtual disk. The interference of PhPs with $\vec{k}_H$ wavevectors yields a convex interference pattern and focal spot (cyan dashed circle) resembling the results obtained in (d).

The numerical study performed above indicates that focusing of in-plane hyperbolic PhPs can be obtained by employing extended optical nanoantennas such as metal disks, i.e. optical elements with convex geometry. Note that this is in stark contrast to the typical concave nanoantennas used to focus polaritons in isotropic media (see inset to Fig. 1d). Such anomalous behavior can be understood by considering the Huygens' principle in hyperbolic media where a circular metal nanoantenna can be seen as an extended source composed of an infinite number of point-like dipoles situated along its edge that launch PhPs whose wavefronts interfere. Figure 1f shows the calculated near field for a discrete number of point electric dipoles placed on a α-MoO$_3$ slab along a semicircle that mimics the nanoantenna's periphery. Contrary to isotropic media where circular fringes parallel to the nanoantenna's periphery are obtained (see Fig. S8 in Supplementary Information), and thus the energy flows equally in all the directions, the interference of highly directional PhPs with $\vec{k}_H$ wavevectors launched by point dipoles leads to focusing into a spot of nanometer dimensions (blue dashed circle in Fig. 1f). This result closely reproduces the near-field distribution obtained for a disk nanoantenna in Fig. 1d, thus confirming that the ray-like polaritons with $\vec{k}_H$ wavevectors along directions closely aligned with the asymptotes of the IFCs are responsible for the foci formation.

**Visualization of the focusing of in-plane hyperbolic PhPs employing metal Au nanoantennas.**
To experimentally demonstrate focusing of in-plane hyperbolic PhPs, we fabricate an Au disk nanoantenna on top of a α-MoO$_3$ slab and perform near-field nanoimaging employing scattering-type scanning near-field optical microscopy (s-SNOM) (see Methods), as sketched in Fig. 2a [22,23]. The images obtained for three different incident wavelengths λ$_0$=10.70, 10.85, and 11.05 μm are shown in Fig. 2b (top panels). All of them reveal a series of convex fringes that emerge from the disk nanoantennas and narrow with the distance with respect to the disk edge, eventually leading to the formation of a focal spot, in excellent agreement with full-wave numerical simulations (bottom panels). Interestingly, we observe that both the width of the focal spot and its focal distance *f* (defined as the distance measured from the focal spot to the nanoantenna's edge along a perpendicular line to the edge) vary as a function of the incident wavelength λ$_0$. By taking profiles along the [001] α-MoO$_3$ direction passing through the focal spots (indicated by arrows in the top panels), we extract spot sizes (corresponding to the FWHM) that vary from ~ 310 nm at λ$_0$ =10.70 μm to ~ 430 nm at λ$_0$ =11.05 μm, which reveal a deep sub-wavelength character, reaching values as small as λ$_0$/34 (Fig. 2c). On the other hand, *f* varies strongly from ~0.6 μm at λ$_0$ =10.70 μm up to ~1.7 μm at λ$_0$ =11.05 μm. This wavelength dependence can be qualitatively understood by the wavelength dependence of the analytically calculated hyperbolic IFC itself (Fig. 2d), which dictates the direction of propagation of PhPs with $\vec{k}_H$ wavevectors that yield the formation of the foci (the Poynting vector $\vec{S}$ denoting the propagation of PhPs with $\vec{k}_H$ wavevectors forms an angle *θ* with respect to the y-axis in Figs. 2b, and 2d). Hence, the particular curvature of the IFC for each of the incident wavelengths λ$_0$ dictates both the angle *θ* and the PhPs wavelengths λ$_H$, being $|\vec{k}_H| = 2\pi/\lambda_H$. Interestingly, whereas the former leads to a wavelength-dependent focal distance *f*, which might find interesting applications in frequency-selective waveguiding, the latter determines the size of the focal spot.

We note that the focal distance *f* can be analytically calculated using a simple equation (1) derived from the Dyadic Green's function where the α-MoO$_3$ crystal is formally treated as a two-dimensional conductive layer (see Supplementary Information):

$$f = R\sqrt{1 - \frac{\varepsilon'_{xx}}{\varepsilon'_{yy}}} \qquad (1)$$

where $\varepsilon'_{xx}$ and $\varepsilon'_{yy}$ are the real part of the α-MoO$_3$ dielectric function for the components "x" and "y", and $R$ is the radius of the disk nanoantenna. By comparing the experimental values of $f$ with those obtained by numerical simulations and analytical calculations, we obtain an excellent agreement (Figs. 2d and 2e). As such, despite its simplicity, equation (1) provides a precise mean for the design of disk nanoantennas capable of focusing hyperbolic PhPs with $\vec{k}_H$ wavevectors at pre-defined distances.

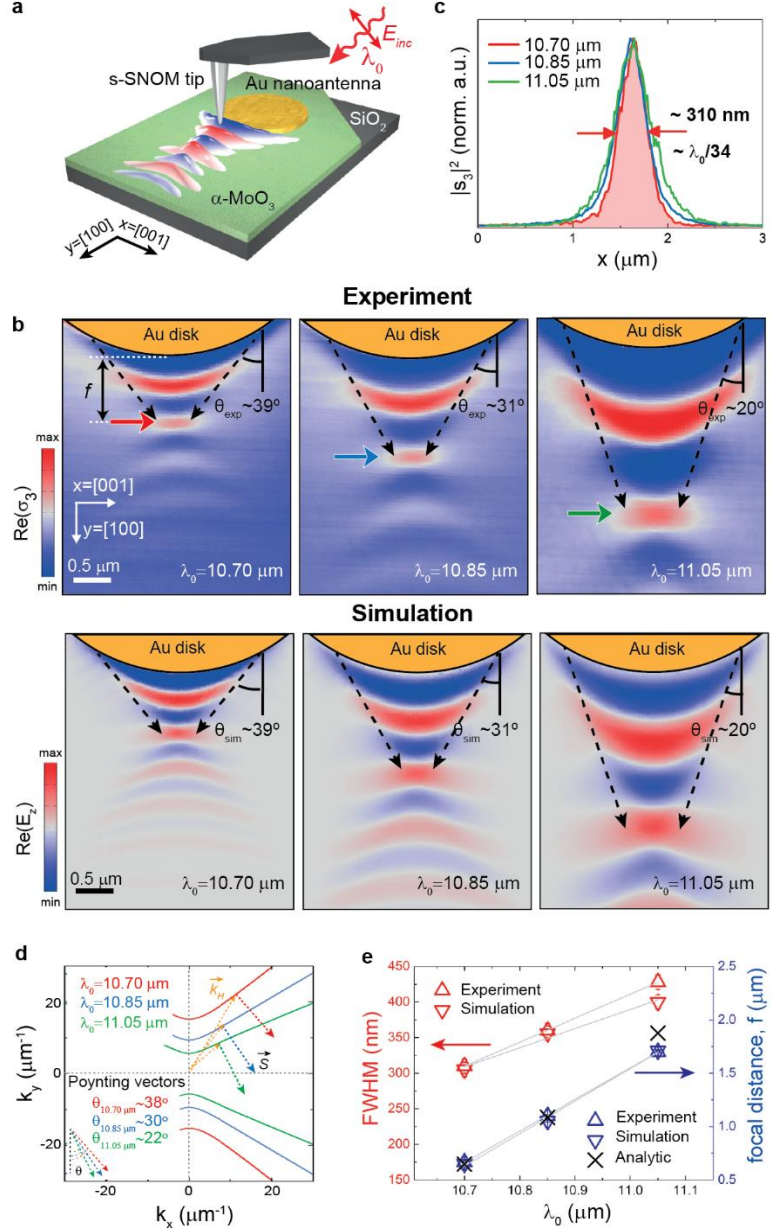

**Figure 2.- Planar focusing of in-plane hyperbolic PhPs with Au disk nanoantennas. a,** Schematics of the s-SNOM experiment to image the propagation and focusing of in-plane hyperbolic PhPs excited by an Au disk nanoantenna on a α-MoO$_3$ slab. The sample is illuminated with p-polarized IR light of wavelength $\lambda_0$. **b,** Experimental ($Re(\sigma_3)$, top row) and simulated ($Re(E_z)$, bottom row) near-field images of PhPs launched by an Au disk nanoantenna fabricated on top of a 165-nm-thick α-MoO$_3$ crystal at $\lambda_0$=10.70 μm (left), $\lambda_0$=10.85 μm (middle) and $\lambda_0$=11.05 μm (right). The interference of PhPs with $\vec{k}_H$ wavevectors launched from the edges of the antenna yields a focal spot with varying size and $f$ as a function of $\lambda_0$. The dashed arrows mark the angle θ. **c,** Experimental near-field amplitude $|s_3|^2$ profiles along the x-axis at positions marked with an arrow in (b) for $\lambda_0$=10.70 μm (red), $\lambda_0$=10.85 μm (blue) and $\lambda_0$=11.05 μm (green). A deep-subwavelength spot size of $\lambda_0/34$ (~310 nm) is measured for $\lambda_0$=10.70 μm. **d,** Analytical IFCs for a 165-nm-thick α-MoO$_3$ crystal at $\lambda_0$=10.70 μm (red), $\lambda_0$=10.85 μm (blue) and $\lambda_0$=11.05 μm (green). The Poynting vector $\vec{S}$ of PhPs with $\vec{k}_H$ wavevectors forms wavelength-dependent angles $\theta$ with respect to the y-axis. **e,** Dependence of the experimental, simulated, and analytically calculated values of the spot size (FWHM) and $f$ with $\lambda_0$ (gray lines serve as a guide for the eye).

**Optimization of the focusing of in-plane hyperbolic PhPs by using rod-like trapezoidal nanoantennas.** At this point, we demonstrated that metal disk nanoantennas allow for planar focusing of hyperbolic PhPs into deeply sub-wavelength spot sizes. However, taking into account that this focusing phenomenon stems from the interference of PhPs with large wavevectors $\vec{k}_H$, which potentially can take infinitely large values, it is reasonable to think that the size of the foci can be further reduced upon optimization of the nanoantennas' design. It should be noted, however, that the minimum size achievable is ultimately conditioned by the extremely short propagation length of PhPs with arbitrarily large wavevectors, thus introducing a trade-off between PhPs wavevector and propagation length. Additionally, the spot size is also affected by the contribution of PhPs with low-$|\vec{k}|$ ($\vec{k}_L$) wavevectors, with relatively larger wavelengths with respect to $\vec{k}_H$ wavevectors.

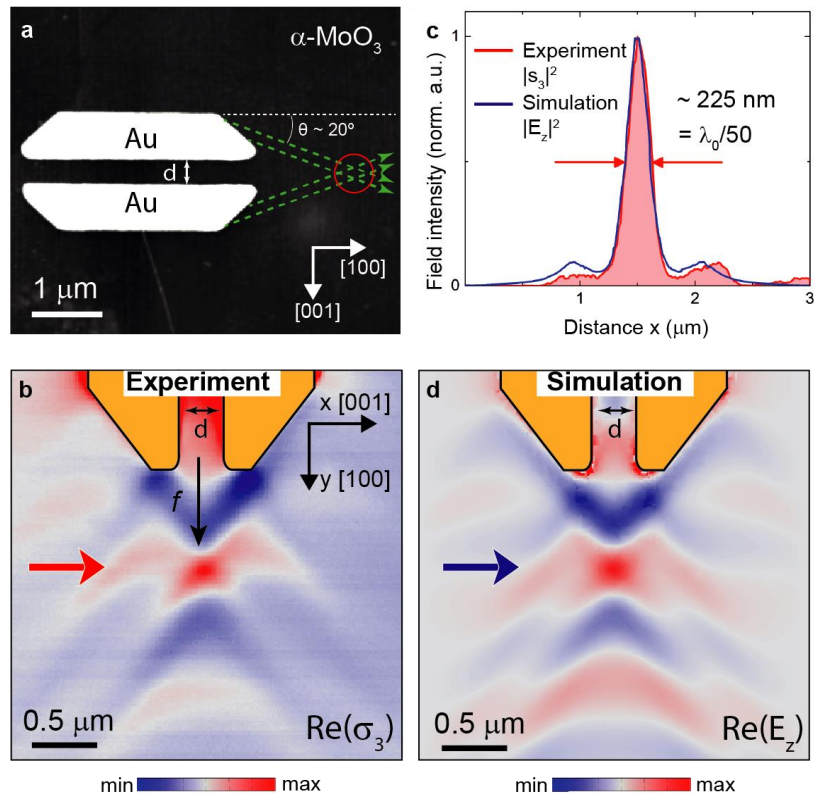

**Figure 3.- Optical nanoantennas for an optimized planar nanofocusing of hyperbolic PhPs. a,** Topographic image of rod-like trapezoidal Au nanoantennas separated by a distance $d$=320 nm on a α-MoO$_3$ crystal. The slope of the antennas' edges at both extremities present an angle of ~44° with respect to the [100] direction. The green dashed arrows illustrate the propagation of PhPs with $\vec{k}_H$ wavevectors excited from the edges of the antennas ($\lambda_0$=11.05 μm) that interfere at the focal spot marked with a red circle. **b,** Experimental near-field image, $Re(\sigma_3)$, of rod-like trapezoidal nanoantennas on a α-MoO$_3$ slab. **c,** Comparison between experimental/simulated near-field amplitude $|s_3|^2/|E_z|^2$ profiles taken along the x-axis ([001] direction) at positions marked by a red/blue arrow in (b) and (d), respectively. A deep sub-wavelength focusing of $\lambda_0/50$ ($\lambda_P/5$) is obtained. **d,** Simulated near-field images, $Re(E_z)$, of rod-like trapezoidal nanoantennas on a α-MoO$_3$ slab.

Bearing this in mind, to further reduce the size of the foci, we consider rod-like trapezoidal nanoantennas in which the extremities are tilted at an angle that depends on $\lambda_0$, taking a value of ~44° for $\lambda_0$=11.05 µm (Fig. 3a). Note that the nanoantenna's geometry needs to be optimized to enhance the near-field intensity and absorption at the focus (see Supplementary Information). In comparison with the disk nanoantennas shown in Fig. 2, this optimal design allows for: *i)* launching PhPs with $\vec{k}_H$ along a unique and well-defined direction (dashed green lines), and *ii)* inhibiting the contribution of PhPs with $\vec{k}_L$ wavevectors to the spot size since the central part of the primitive disk-like nanoantenna is absent, i.e., only ray-like PhPs with $\vec{k}_H$ wavevectors are contributing to the focal spot. Figure 3b shows the experimental near-field image, $Re(\sigma_3)$, for a rod-like trapezoidal nanoantenna fabricated on a α-MoO$_3$ slab. We observe PhPs with $\vec{k}_H$ wavevectors being excited at the edges of the nanoantennas which upon propagation interfere, giving rise to the formation of a focal spot at a distance $f$ ~790 nm. By taking a profile along the x-axis ([001] α-MoO$_3$ direction) passing through the focus (red curve in Fig. 3c), we extract a spot size (FWHM) as small as ~225 nm, i.e., a deep sub-wavelength size of ~$\lambda_0$/50, or ~$\lambda_p$/5, being $\lambda_p$~1.2 µm the polaritonic wavelength along the y-axis ([100] α-MoO$_3$ direction). This result is in excellent agreement with the value obtained from numerical simulations (blue curve in Fig. 3c corresponding to a profile taken through the focal spot in the simulated near-field image of Fig. 3d). We note that such as in the case of the disk-like nanoantenna, the focal distance $f$ for rod-like trapezoidal nanoantennas can also be analytically derived using a simple equation (equation S28 in Supplementary Information), which gives a value $f$~760 nm, in excellent agreement with the experiment and numerical simulation ($f$ ~790 nm).

The results above demonstrate that by simply tailoring the geometry of Au nanoantennas with a trapezoidal shape, allowing only the interference of polaritons with enhanced density of electromagnetic modes, the near-field confinement at the focus position can be enhanced. Furthermore, the focal distance $f$ of these nanoantennas can be easily tuned by simply varying the separation distance "$d$" between the trapezoidal rods, thus providing an additional tuning knob for a fixed illuminating wavelength (see Supplementary Information).

**Enhanced power dissipation and near-field intensity at the focus.** Focusing of light into deeply sub-wavelength regions at the nanoscale has profound implications for fundamental studies on near-field light-matter interactions, energy harvesting or heat management [24–27]. To assess the potential of our hyperbolic foci for these applications, we calculate both the power dissipation ($P_{dis}$), and the electric field intensity ($I$) at the focus location (Methods) and compare them to the case of foci in isotropic media (Methods). Specifically, we perform full-wave numerical simulations and integrate $P_{dis}$ and $I$ along a horizontal profile at the focal spot positions for the case of hyperbolic media: using both rod-like trapezoidal nanoantennas (Fig. 4a) and a disk nanoantenna (Fig. 4b), and the case of isotropic media: using a typical rod-like concave nanoantenna (Fig. 4c). In all cases, the geometries of the nanoantennas are adjusted to guarantee the same focal distance $f$ ~ 1.7 µm. For a more comprehensive comparison, we also consider the PhPs wavelength along the y-axis in Fig. 4 to be equal in both media. Interestingly, we find that the focusing of PhPs in hyperbolic media leads to a markedly enhanced near-field intensity, near-field confinement, and power dissipation at the focal spot, as depicted in Fig. 4d and 4e, respectively. Particularly, in the case of a disk-like nanoantenna, the near-field intensity $|E|^2$ and integrated $P_{dis}$ are enhanced with respect to the isotropic case up to a factor of x2 and x4, respectively. Note that for a reasonable comparison between the performance of the nanoantennas in hyperbolic and isotropic media, the power dissipation and near-field intensity magnitudes are normalized to the field inside and above the nanoantennas, respectively (see Methods). Besides, the focus FWHM is dramatically decreased from ~600 nm, i.e., $\lambda_p$/2 (isotropic case using rod-like nanoantennas with concave extremities) down to ~225 nm, i.e., $\lambda_p$/5 (hyperbolic case using rod-like trapezoidal nanoantennas). Altogether, these results reveal that rod-like trapezoidal nanoantennas give rise to foci in hyperbolic media with 2.5 times larger near-field confinement than that obtained in isotropic media with conventional concave antennas. These results suggest directional in-plane propagation of hyperbolic PhPs with $\vec{k}_H$ wavevectors as a key characteristic that offers unprecedented possibilities in nanophotonics. For example, focusing of such PhPs to deeply subwavelength volumes would represent an extraordinary resource for harvesting of IR light and enhancing light-matter interactions [18,28].

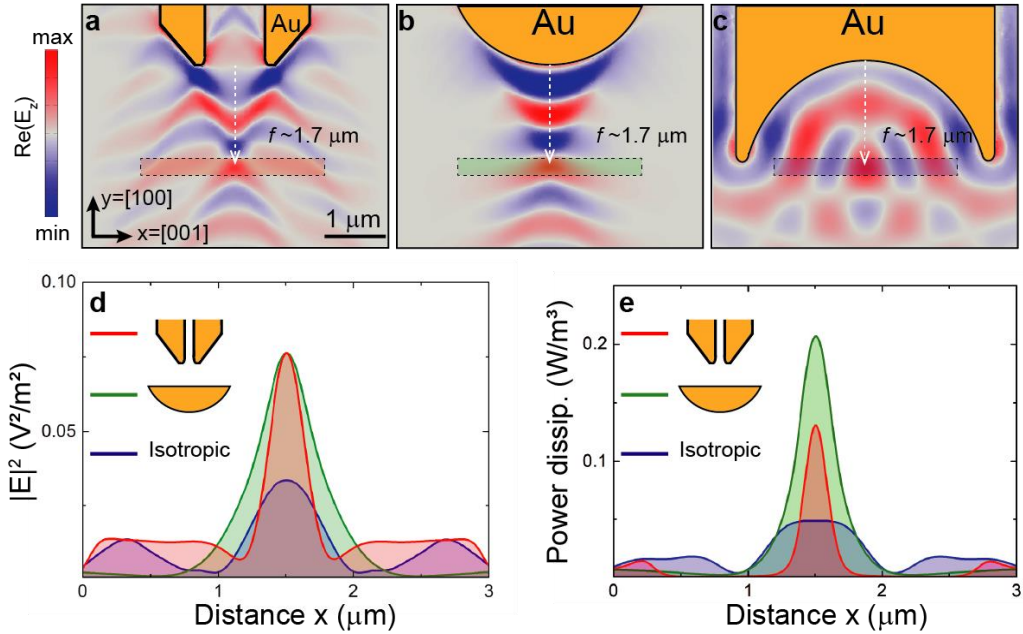

**Figure 4.- In-plane hyperbolic foci reveal enhanced near-field intensity, field confinement and power dissipation when compared to foci in isotropic media.** Simulated near-field maps, $Re(E_z)$, showing the propagation and focusing of PhPs on the surface of a 165-nm-thick α-MoO$_3$ crystal for an illuminating wavelength $\lambda_0$=11.05 μm employing: **a,** rod-like trapezoidal nanoantennas; **b,** disk-like nanoantenna. **c,** rod-like Au nanoantenna with concave extremities in an isotropic medium. **d,** Integrated near-field intensity, $|E|^2$, above the surface of the slab along the red/green/blue shadowed volume in (a, b and c), respectively. The data are normalized to the integrated $|E|^2$ value above the nanoantennas (see Methods). **e,** Integrated power dissipation density $P_{dis}$ within the slab along the red/green/blue shadowed volume in (a, b and c), respectively. The data are normalized to the integrated power dissipation density value in the nanoantennas (see Methods).

**Conclusions.** We demonstrate the possibility of exciting and focusing highly directional PhPs with $\vec{k}_H$ wavevectors along the surface of a hyperbolic medium (α-MoO$_3$) by employing optical metal nanoantennas with tailored geometries, which thus constitute the first nano-optical elements for planar nano-optics in hyperbolic media. Interestingly, the enhanced near-field intensity, field confinement and power density dissipation provided by focusing in-plane hyperbolic PhPs might offer exciting possibilities for applications in nanophotonics such as infrared frequency-selective waveguiding for nanoscale spectrometry, light routing, light-matter interaction experiments as well as heat management. Based on these findings, we envision a planar nanophotonics field where the joint advantages of enhanced near-field confinement and energy management at the nanoscale in strongly anisotropic media, together with tunability by strain fields, electric gating or near-field hybridization in vdW heterostructures, might open the door for more efficient applications in bio-chemical sensing or near-field thermal harvesting.

**REFERENCES**

1. Schnell, M. et al. Nanofocusing of mid-infrared energy with tapered transmission lines. Nat. Photonics 5, 283-287 (2011)

2. Kravtsov, V., Ulbricht, R., Atkin, J. M. & Raschke, M. B. Plasmonic nanofocused four-wave mixing for femtosecond near-field imaging. Nat. Nanotechnol. 11, 459-464 (2016)
3. Tsakmakidis, K. L., Hess, O., Boyd, R. W. & Zhang, X. Ultraslow waves on the nanoscale. Science 358, eaan5196 (2017)
4. Luo, X. & Ishihara, T. Surface plasmon resonant interference nanolithography technique. Appl. Phys. Lett. 84, 4780(2004)
5. Ishii, S., Kildishev, A. V., Narimanov, E., Shalaev, V. M. & Drachev, V. P. Sub-wavelength interference pattern from volume plasmon polaritons in a hyperbolic medium. Laser Photonics Rev. 7, 265–271 (2013).
6. Kim, Y., Smith, J. G. & Jain, P. K. Harvesting multiple electron-hole pairs generated through plasmonic excitation of Au nanoparticles. Nat. Chem. 10, 763-769 (2018)
7. Christopher, P., Xin, H. & Linic, S. Visible-light-enhanced catalytic oxidation reactions on plasmonic silver nanostructures. Nat. Chem. 3, 467-472 (2011) doi:10.1038/nchem.1032.
8. Rodrigo, D. et al. Mid-infrared plasmonic biosensing with graphene. Science 349, 165–168 (2015).
9. Autore, M. et al. Boron nitride nanoresonators for Phonon-Enhanced molecular vibrational spectroscopy at the strong coupling limit. Light Sci. Appl. 7, 17172 (2018)
10. Nie, S. & Emory, S. R. Probing single molecules and single nanoparticles by surface-enhanced Raman scattering. Science 275, 1102-1106 (1997)
11. Basov, D. N., Fogler, M. M. & García De Abajo, F. J. Polaritons in van der Waals materials. Science 354, aag1992 (2016)
12. Xu, X. G. et al. One-dimensional surface phonon polaritons in boron nitride nanotubes. Nat. Commun. 5, 1–6 (2014).
13. Dai, S. et al. Tunable Phonon Polaritons in Atomically Thin van der Waals Crystals of Boron Nitride. Science 343, 1125-1129 (2014).
14. Li, P. et al. Optical nanoimaging of hyperbolic surface polaritons at the edges of van der Waals materials. Nano Lett. 17, 228–235 (2017).
15. Li, P. et al. Hyperbolic phonon-polaritons in boron nitride for near-field optical imaging and focusing. Nat. Commun. 6, 7507 (2015)
16. Dai, S. et al. Subdiffractional focusing and guiding of polaritonic rays in a natural hyperbolic material. Nat. Commun. 6, 6963 (2015)
17. Nikitin, A. Y. et al. Nanofocusing of Hyperbolic Phonon Polaritons in a Tapered Boron Nitride Slab. ACS Photonics 3, 924-929 (2016)
18. Kildishev, A. V., Boltasseva, A. & Shalaev, V. M. Planar photonics with metasurfaces. Science 339, 1232009 (2013)
19. Li, P. et al. Infrared hyperbolic metasurface based on nanostructured van der Waals materials. Science 359, 892-896 (2018)
20. Taboada-Gutiérrez, J. et al. Broad spectral tuning of ultra-low-loss polaritons in a van der Waals crystal by intercalation. Nat. Mater. 19, 964–968 (2020)
21. Ma, W. et al. In-plane anisotropic and ultra-low-loss polaritons in a natural van der Waals crystal. Nature 562, 557–562 (2018)
22. Chen, X. et al. Modern Scattering-Type Scanning Near-Field Optical Microscopy for Advanced Material Research. Advanced Materials 31, 1804774 (2019)
23. Hillenbrand, R., Taubner, T. & Keilmann, F. Phonon-enhanced light-matter interaction at the nanometre scale. Nature 418, 159–162 (2002)
24. Barnes, W. L., Dereux, A. & Ebbesen, T. W. Surface plasmon subwavelength optics. Nature 424, 824–830 (2003)
25. Gramotnev, D. K. & Bozhevolnyi, S. I. Nanofocusing of electromagnetic radiation. Nature Photonics 8, 13–22 (2014)
26. Kim, S. et al. High external-efficiency nanofocusing for lens-free near-field optical nanoscopy. Nat. Photonics 13, 636–643 (2019)
27. Boriskina, S. V., Weinstein, L. A., Tong, J. K., Hsu, W. C. & Chen, G. Hybrid Optical-Thermal Antennas for Enhanced Light Focusing and Local Temperature Control. ACS Photonics 3, 1714–1722 (2016)

28. Gomez-Diaz, J. S. & Alù, A. Flatland Optics with Hyperbolic Metasurfaces. ACS Photonics 3, 2211–2224 (2016)

29. Pons-Valencia, P. et al. Launching of hyperbolic phonon-polaritons in h-BN slabs by resonant metal plasmonic antennas. Nat. Commun. 10, 3242 (2019)

30. Fei, Z. et al. Gate-tuning of graphene plasmons revealed by infrared nano-imaging. Nature 487, 82–85 (2012)

31. Chen, J. et al. Optical nano-imaging of gate-tunable graphene plasmons. Nature 487, 77–81 (2012).

32. Álvarez-Pérez, G. et al. Infrared permittivity of the biaxial van der Waals semiconductor α-$MoO_3$ from near- and far-field correlative studies. Adv. Mater. 32, 1908176 (2020)